\documentclass[9pt]{article}
\usepackage{graphicx}
\usepackage{overpic}

\newcommand\pubnumber{DPF2015-19}
\newcommand\pubdate{\today}

\def\napoli{
Institute of High Energy Physics, Beijing, China}

\def\Title#1{\begin{center} {\Large #1 } \end{center}}
\def\Author#1{\begin{center}{ \sc #1} \end{center}}
\def\Address#1{\begin{center}{ \it #1} \end{center}}

\newcommand\pubblock{\rightline{\begin{tabular}{l} \pubnumber\\
         \pubdate  \end{tabular}}}
\newenvironment{Abstract}{\begin{quotation}  }{\end{quotation}}
\newenvironment{Presented}{\begin{quotation} \begin{center}
             PRESENTED AT\end{center}\bigskip
      \begin{center}\begin{large}}{\end{large}\end{center} \end{quotation}}

\def\beq{\begin{equation}}
\def\eeq#1{\label{#1}\end{equation}}
\def\eeqn{\end{equation}}

\def\beqa{\begin{eqnarray}}
\def\eeqa#1{\label{#1}\end{eqnarray}}
\def\eeqan{\end{eqnarray}}

\let\bar=\overbar

\def\Dslash{\not{\hbox{\kern-4pt $D$}}}
\def\dslash{\not{\hbox{\kern-2pt $\del$}}}

\def\msb{{\bar{\ssstyle M \kern -1pt S}}}

\begin{document}
\begin{titlepage}
\pubblock

\vfill
\Title{Exotic and Charmonium(-like) states at BESIII}
\vfill
\Author{ Peilian Liu, on behalf of the BESIII Collaboration}
\Address{\napoli}
\vfill
\begin{Abstract}
The BESIII experiment at the Beijing Electron Positron Collider (BEPCII) has accumulated the world's largest samples of direct $e^+e^-$ collisions in the $\tau$-charm region. From the collected samples, which include $e^+e^-$ annihilations at $J/\psi$, $\psi'$, $\psi$(3770) peaks and in the region from 3.8\,GeV to 4.6\,GeV, BESIII has produced many new physics results in the spectroscopy, transitions, and decays of charmonium(-like) states. This talk will cover the latest results over a wide range of topics from radiative and hadronic transitions among charmonium states, as well as the productions and decays of the $XYZ$ states.
\end{Abstract}
\vfill
\begin{Presented}
DPF 2015\\
The Meeting of the American Physical Society\\
Division of Particles and Fields\\
Ann Arbor, Michigan, August 4--8, 2015\\
\end{Presented}
\vfill
\end{titlepage}

\section{Introduction}
The charmonium particles which contain a charm and an anti-charm quark, has been an excellent tool for probing Quantum Chromodynamics (QCD), the fundamental theory that describes the strong interactions between
quarks and gluons, in the non-perturbative (low-energy, long-distance effects) regime, and remains of high interest both experimentally and theoretically. All of the charmonium states with masses that are below the open-charm threshold have been firmly established~\cite{In1,In2}; open-charm refers to mesons containing a charm quark (antiquark) and either an up or down antiquark (quark), such as $D$ or $\bar{D}$. However, the observation of the spectrum that are above the opencharm threshold remains unsettled.
QCD predicts the new particles beyond the quark model, \emph{e.g.}, glueballs, hybrids and multiquark states.
Experimental study of the decay modes of known charmonium states and the search of new charmonium states would provide helpful information for further verifying the QCD theory.
During the past decade, many new charmonium-like states were discovered, such as the $X$(3872), the $Y$(4260) and the $Z_c$(3900). These states provide strong evidence for the existence of exotic hadron states~\cite{In8}. Although charged charmonium-like states like the $Z_{c}$(3900) provide convincing evidence for the existence of multi-quark states~\cite{In9}, it is more difficult to distinguish neutral candidate exotic states from conventional charmonium states. Moreover, the study of transitions between charmonium-like states,
is an important approach to probe their nature, and the connections between them. Thus, a more complete understanding of the charmonium(-like) spectroscopy and their relations is necessary and timely.

Since the upgrade was completed in 2008, the BESIII~\cite{BES3} detector at the BEPCII collider, has collected the
world's largest data samples at $\tau$-charm energy region, including 1.3 billion $J/\psi$ events, 0.6 billion $\psi'$ events, 2.9\,fb$^{-1}$ at the peak of the $\psi$(3770) resonance, and lots of data samples ranging from 3.8 to 4.6\,GeV, which offers us a unique opportunity to study the charmonium(-like) physics.

\section{$X$ states}
$X(3872)$ is observed by Belle~\cite{belle:3872} for the first time in the process $B^{\pm}\rightarrow K^{\pm}\pi^{+}\pi^{-}J/\psi$ in 2003, which is the first observed exotic state. The mass of this state was determined to be $M=(3872.0\pm0.6\pm0.5)$\,MeV, which is higher than potential model expectations for the center-of-gravity (cog) of the $1^3$D$_{cJ}$ charmonium states~\cite{cog_1,cog_2,cog_3}. The decay of the $^3$D$_{c2}$ charmonium state to $\gamma \chi_{c1}$ is an allowed $E1$ transition with a partial width that is expected to be substantially larger than that for the $\pi^{+}\pi^{-}J/\psi$ final state; e.g. the Ref.~\cite{3Dc2} predicts $\Gamma(^3$D$_{c2}\rightarrow \gamma \chi_{c1})>5\times \Gamma(^3$D$_{c2}\rightarrow \pi^{+}\pi^{-}J/\psi)$. However, based on Belle's results, the ratio of partial widths was determined to be less than 0.89 at a 90\% confidence level (C.L.), which contradicts expectations for the $^3$D$_{c2}$ charmonium state. So $X(3872)$ is a new state, which is not expected but discovered. Because of the proximity of its mass to the $D\bar{D}^*$ mass threshold, the $X(3872)$ has been interpreted as a candidate for a hadronic molecule or a tetraquark state~\cite{X3872-moce-tetraq}.

\subsection{Observation of $e^{+}e^{-}\rightarrow \gamma X(3872)$}
\begin{figure}[!t]
\begin{center}
\includegraphics[width=0.32\textwidth]{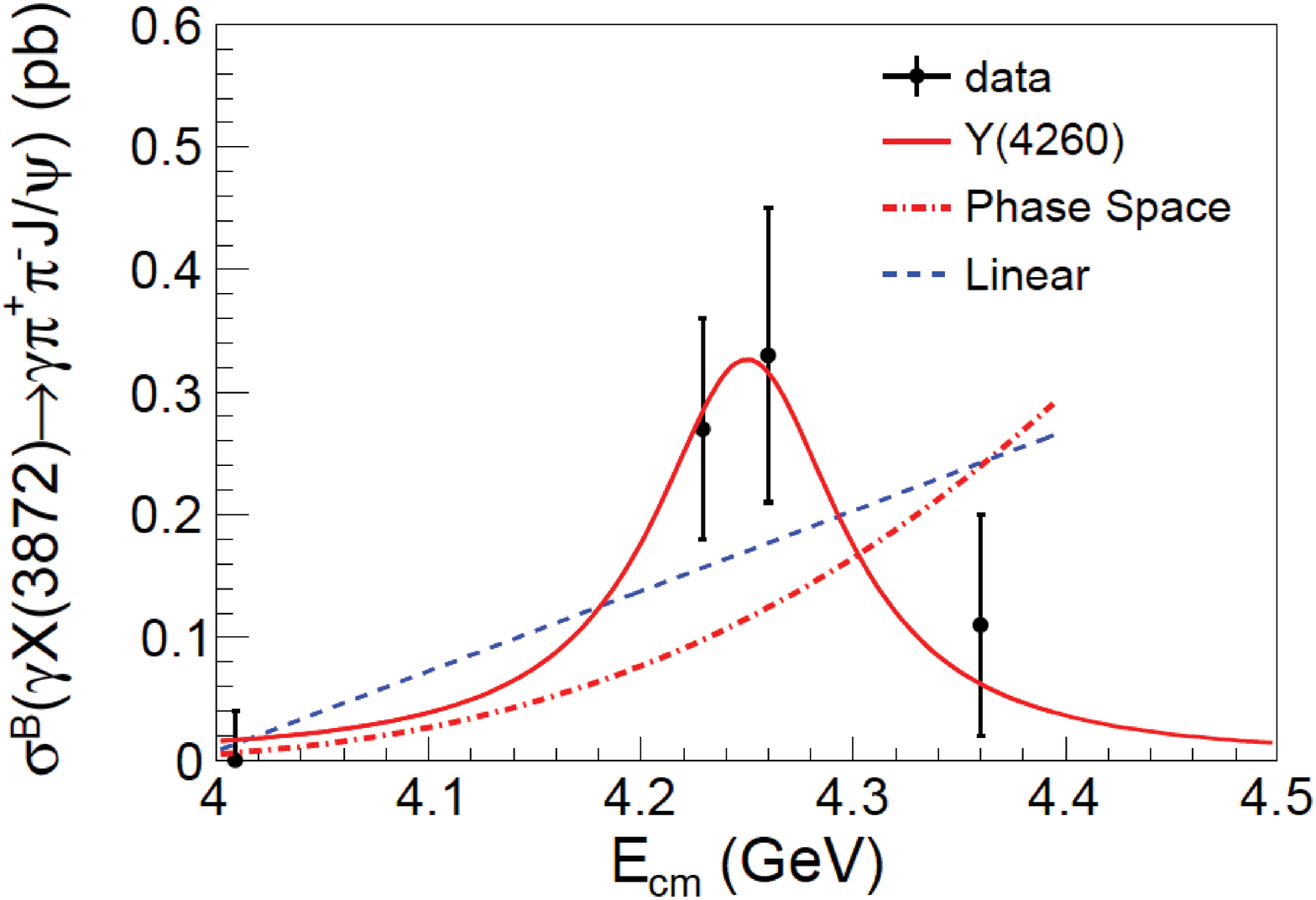}
\includegraphics[width=0.32\textwidth]{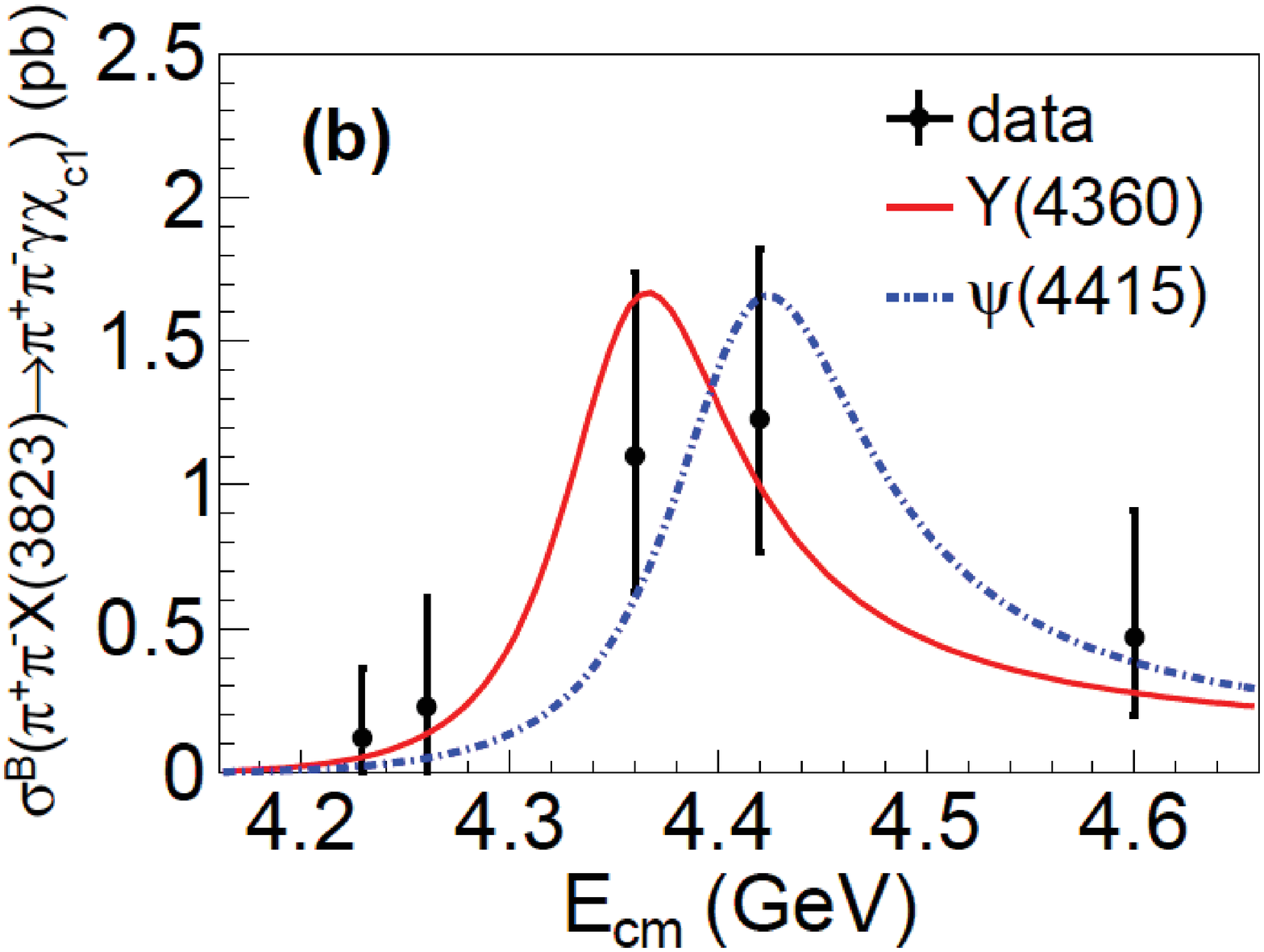}
\includegraphics[width=0.32\textwidth]{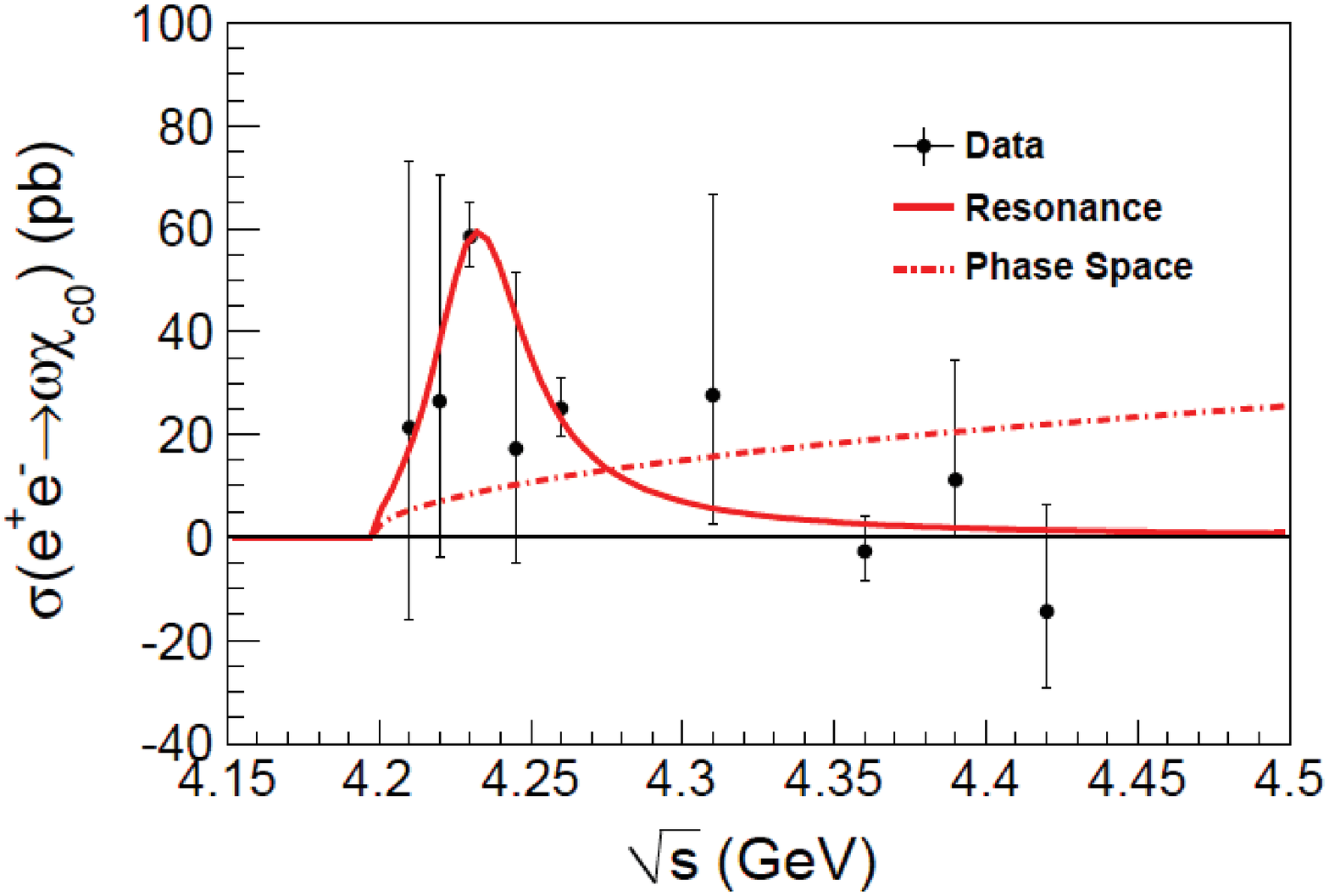}
\caption{(a) The fit to $\sigma^{B}[e^{+}e^{-}\rightarrow \gamma X(3872)]\cdot \mathcal{B}[X(3872)\rightarrow \pi^{+}\pi^{-} J/\psi]$. (b) The fit to $\sigma^B[e^+e^-\rightarrow \pi^+\pi^-X(3823)]\cdot\mathcal{B}(X(3823)\rightarrow\gamma\chi_{c1})$. (c) Fit to $\sigma^B(e^+e^-\rightarrow \omega\chi_{c0})$ with a resonance, or a phase space term.}
\label{fig:X3872-pipijpsi}
\end{center}
\end{figure}

Since the $X(3872)$ is a $1^{++}$ state, it should be able to be produced through the radiative transition of an excited vector charmonium or charmonium-like states such as a $\psi$ or a $Y$.
BESIII reported the first observation of the process $e^{+}e^{-}\rightarrow \gamma X(3872)\rightarrow \gamma\pi^{+}\pi^{-} J/\psi, J/\psi\rightarrow l^{+}l^{-} (l=e,\mu)$.
The measured mass of $X(3872)$ is ($3871.9\pm0.7\pm0.2$)\,MeV/$c^2$.
From a fit with a floating width we obtain $\Gamma[X(3872)]=(0.0^{+1.7}_{-0.0})$\,MeV, or less than 2.4\,MeV at the 90\% C.L.. The statistical significance of $X(3872)$ is 6.3$\sigma$.
The product of the Born cross section times the branching fraction of $X(3872)\rightarrow \pi^{+}\pi^{-} J/\psi$ are shown in Figure~\ref{fig:X3872-pipijpsi} (a), which indicates that the $Y$(4260) resonance describes the data better than the other two options.
Combining with the $e^{+}e^{-}\rightarrow \pi^{+}\pi^{-} J/\psi$ cross section measurement at $\sqrt{s}=4.26$\,GeV from BESIII~\cite{BESpipijpsi}, we obtain $\sigma^{B}[e^{+}e^{-}\rightarrow \gamma X(3872)]\cdot \mathcal{B}[X(3872)\rightarrow \pi^{+}\pi^{-} J/\psi]$/$\sigma^{B}(e^{+}e^{-}\rightarrow \pi^{+}\pi^{-} J/\psi)$ = $(5.2\pm1.9)\times10^{-3}$. If we take $\mathcal{B}[X(3872)\rightarrow \pi^{+}\pi^{-} J/\psi] = 5$\%, we get $\frac{\mathcal{B}(Y(4260)\rightarrow \gamma X(3872))}{\mathcal{B}(Y(4260)\rightarrow \pi^{+}\pi^{-} J/\psi)}$ = 0.1.

\subsection{Observation of $X(3823)$ state in $e^{+}e^{-}\rightarrow \pi^{+}\pi^{-}\gamma\chi_{c1}$}
BESIII observed a narrow resonance, $X(3823)$, through the process $e^+e^-\rightarrow \pi^+\pi^-X(3823)$ with a statistical significance of 6.2$\sigma$. The measured mass of the $X(3823)$ is $(3821.7\pm1.3)$\,MeV/$c^2$, and the width is less than 16\,MeV at the 90\% C.L.
The energy-dependent cross sections of $\sigma^B(e^+e^-\rightarrow \pi^+\pi^-X(3823))\cdot\mathcal{B}(X(3823)\rightarrow\gamma\chi_{c1})$ are fitted in Figure~\ref{fig:X3872-pipijpsi} (b), which shows that both the $Y$(4360) and the $\psi$(4415) hypotheses are acceptable.
The $X(3823)$ resonance is a good candidate for the $\psi(1^3D_2)$ charmonium state. Since their masses are consistent. And our measured ratio $\frac{\mathcal{B}(X(3823)\rightarrow\gamma\chi_{c2})}{\mathcal{B}(X(3823)\rightarrow\gamma\chi_{c1})}<0.42$ (at the 90\% C.L.) agrees with expectations for the $\psi(1^3D_2)$ state.

\section{$Y$ states}
During the last decade, new charmonium-like vector states, such as the $Y$(4260), $Y$(4360) and $Y$(4660), have been observed by BABAR~\cite{y1,y2}, Belle~\cite{y3,y4,y5,y6} and CLEO~\cite{y7}. The masses of these new $Y$ states are above the $D\bar{D}$ production threshold, ranging from 4.0 to 4.7\,GeV/$c^2$. Since all of them are produced in $e^+e^-$ annihilation, and since they have been observed to decay in dipion hadronic transitions to the $J/\psi$ or $\psi$(3686), one would naturally interpret these states as vector charmonium excitations. However, peculiar features of these $Y$ states reveal an exotic nature that likely excludes a conventional charmonium interpretation. These features include a discrepancy with the spectrum of vector charmonium states predicted by the potential model, a surprisingly large coupling to final states without open-charm mesons, and a lack of observation in the inclusive hadronic cross section.
Searching for new decay modes and measuring the line shapes of their production cross sections will be very helpful for these $Y$ states interpretation. Hadronic transitions (by $\eta$, $\pi^0$, or a pion pair) to lower charmonia like the $J/\psi$ are also regarded as sensitive probes to study the properties of these $Y$ states~\cite{y22}.

\subsection{Cross sections of $e^+e^-\rightarrow\pi\pi J/\psi(h_{c})$}~\label{sec.pipijpsi/hc}
\begin{figure}[!htb]
\begin{center}
\includegraphics[width=0.32\textwidth]{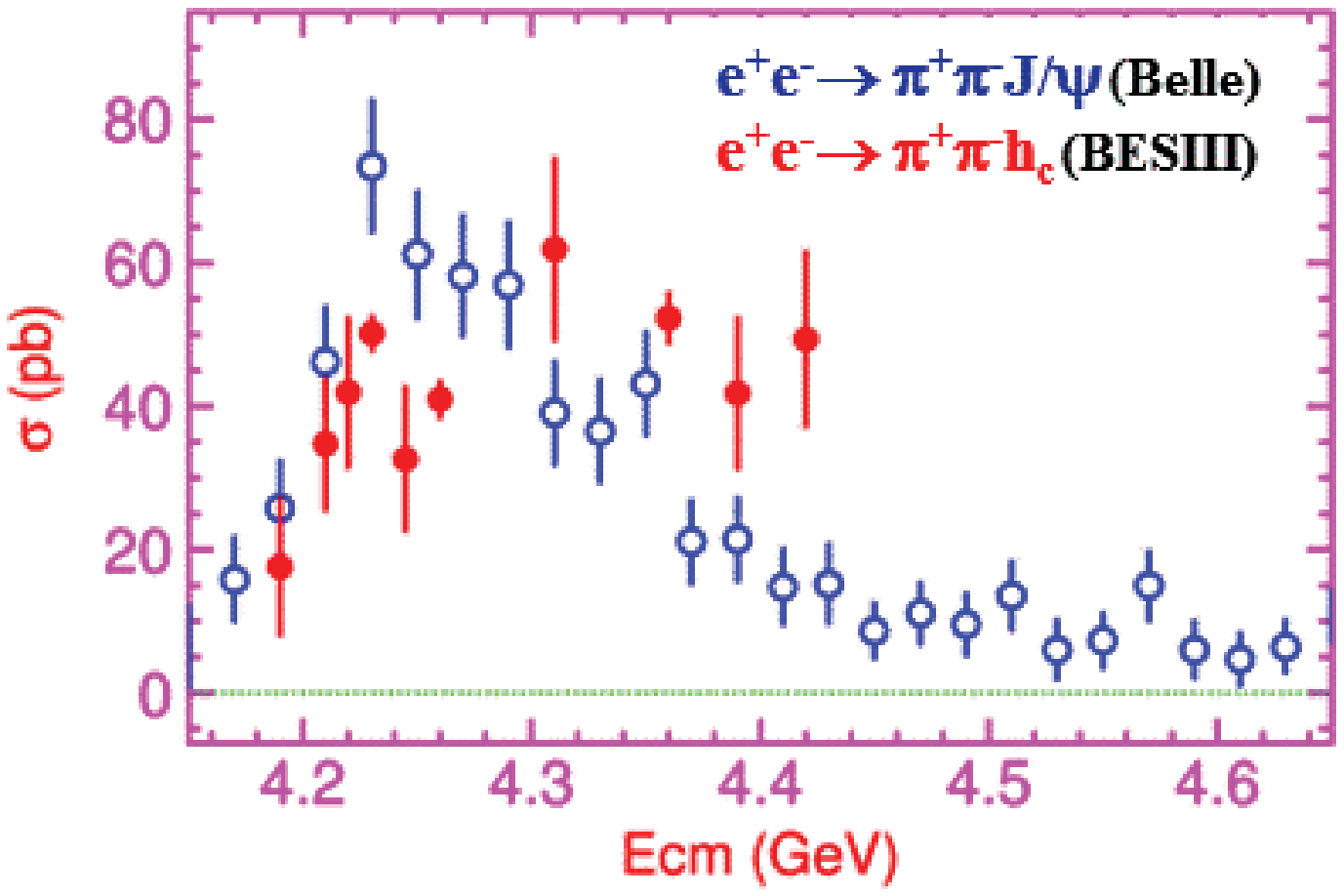}
\includegraphics[width=0.32\textwidth]{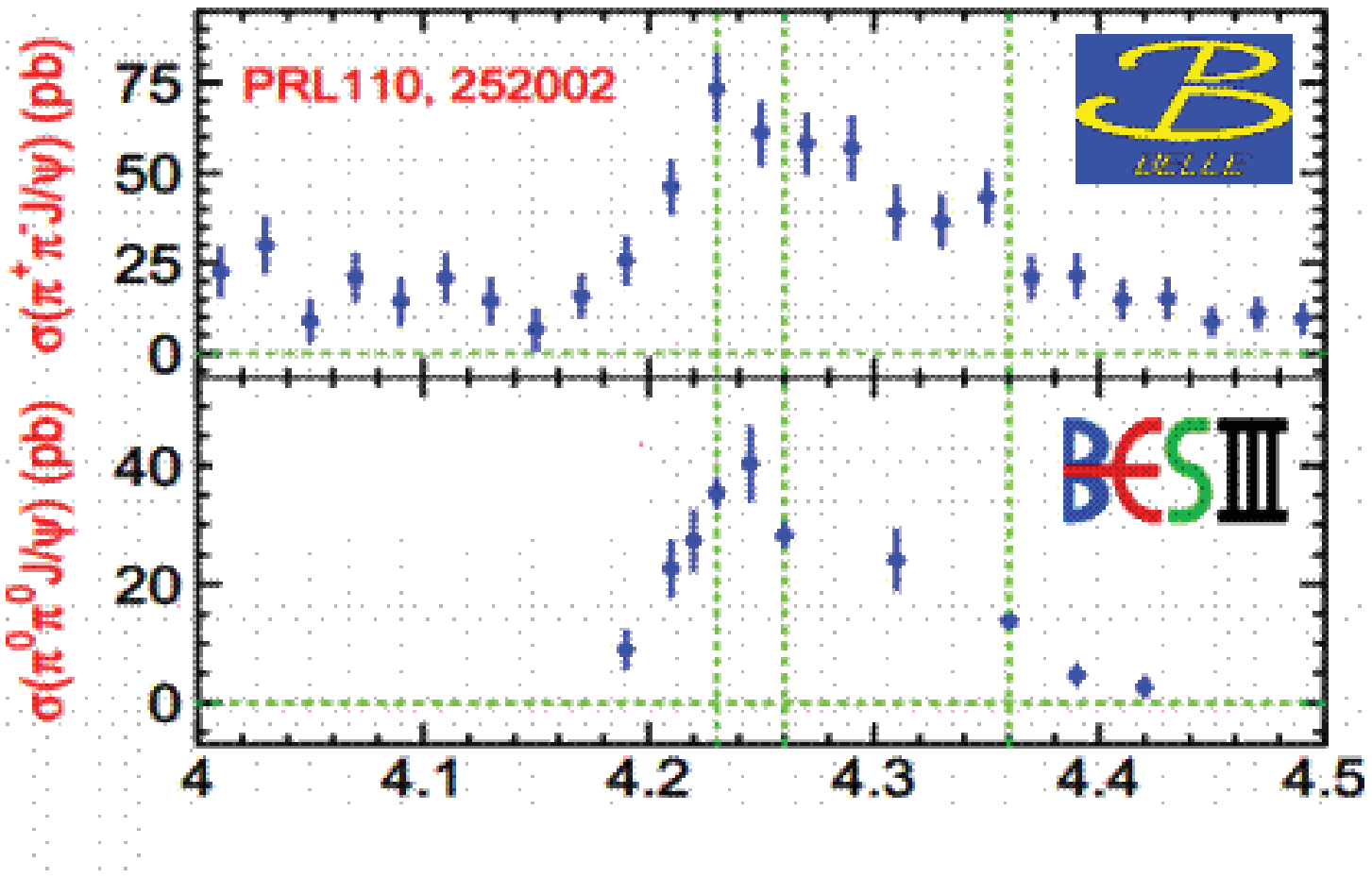}
\includegraphics[width=0.32\textwidth]{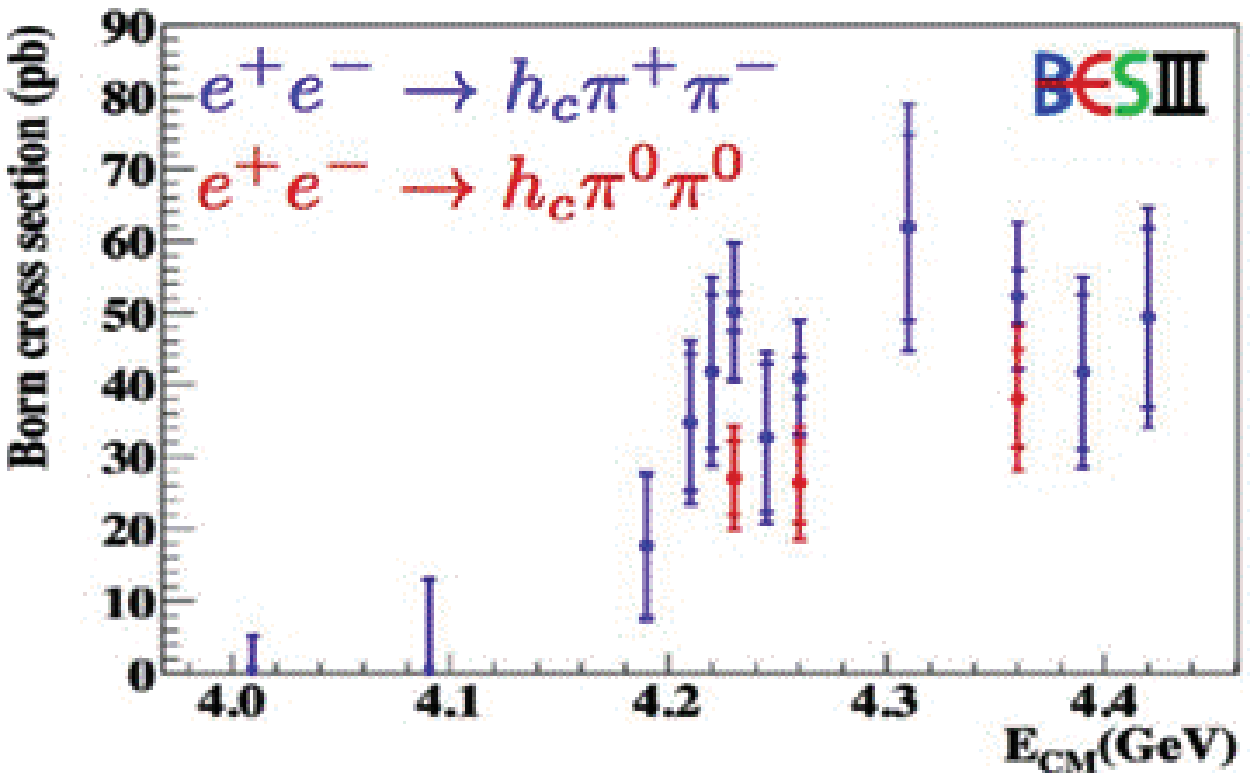}
\caption{Cross sections of $e^+e^-\rightarrow\pi\pi J/\psi$ and $e^+e^-\rightarrow\pi\pi h_{c}$.}
\label{fig:pipijpsi/hc}
\end{center}
\end{figure}

BESIII measured the cross sections of $e^+e^-\rightarrow\pi^+\pi^-h_{c}$~\cite{pippimhc}, $e^+e^-\rightarrow\pi^0\pi^0J/\psi$~\cite{pi0pi0jpsi} and $e^+e^-\rightarrow\pi^0\pi^0h_{c}$~\cite{pi0pi0hc} for the first time.
As shown in Figure~\ref{fig:pipijpsi/hc} (a), the cross sections of $e^+e^-\rightarrow\pi^+\pi^-h_{c}$ are of the same order of magnitude as those of the $e^+e^-\rightarrow\pi^+\pi^-J/\psi$ measured by Belle~\cite{y4}, but with a different line shape. In Figure~\ref{fig:pipijpsi/hc} (b), the measured Born cross sections of $e^+e^-\rightarrow\pi^0\pi^0J/\psi$ are about half of those for $e^+e^-\rightarrow\pi^+\pi^-J/\psi$ that were measured by Belle, consistent with the isospin symmetry expectation for resonances. The Born cross sections of $e^+e^-\rightarrow\pi^0\pi^0h_{c}$ are about half of those of $e^+e^-\rightarrow\pi^+\pi^-h_{c}$ within less than $2\sigma$, in Figure~\ref{fig:pipijpsi/hc} (c). So, no distinct isospin violation are observed in the cross sections for both $e^+e^-\rightarrow \pi\pi J/\psi$ and $e^+e^-\rightarrow \pi\pi h_{c}$.

\subsection{Cross sections of $e^+e^-\rightarrow \omega\chi_{c0}$}
BESIII reported the first observation of $e^+e^-\rightarrow\omega\chi_{c0}$, whilst no obvious signals of $e^+e^-\rightarrow \omega\chi_{c1,c2}$ are observed. The cross sections for $e^+e^-\rightarrow \omega\chi_{c0}$ are shown in Figure~\ref{fig:X3872-pipijpsi} (c) which indicates that the resonance describes the data better than the phase space term. The mass and width of the resonance are determined to be $(4230\pm8\pm6)$\,MeV/$c^2$ and $(38\pm12\pm2)$\,MeV, respectively. The parameters are inconsistent with those obtained by fitting a single resonance to the $\pi^+\pi^-J/\psi$ cross section~\cite{babar-2005}. This suggests that the observed $\omega\chi_{c0}$ signals be unlikely to originate from the $Y(4260)$.

Since the cross sections of $e^+e^-\rightarrow\pi^+\pi^-h_{c}$ and $e^+e^-\rightarrow \omega\chi_{c0}$ are inconsistent with the line shape of $e^+e^-\rightarrow\pi^+\pi^-J/\psi$, this hints at the existence of a more complicated and mysterious underlying dynamics.

\subsection{Cross sections of $e^+e^-\rightarrow \eta J/\psi$}
\begin{figure}[!htb]
\begin{center}
\includegraphics[width=0.32\textwidth]{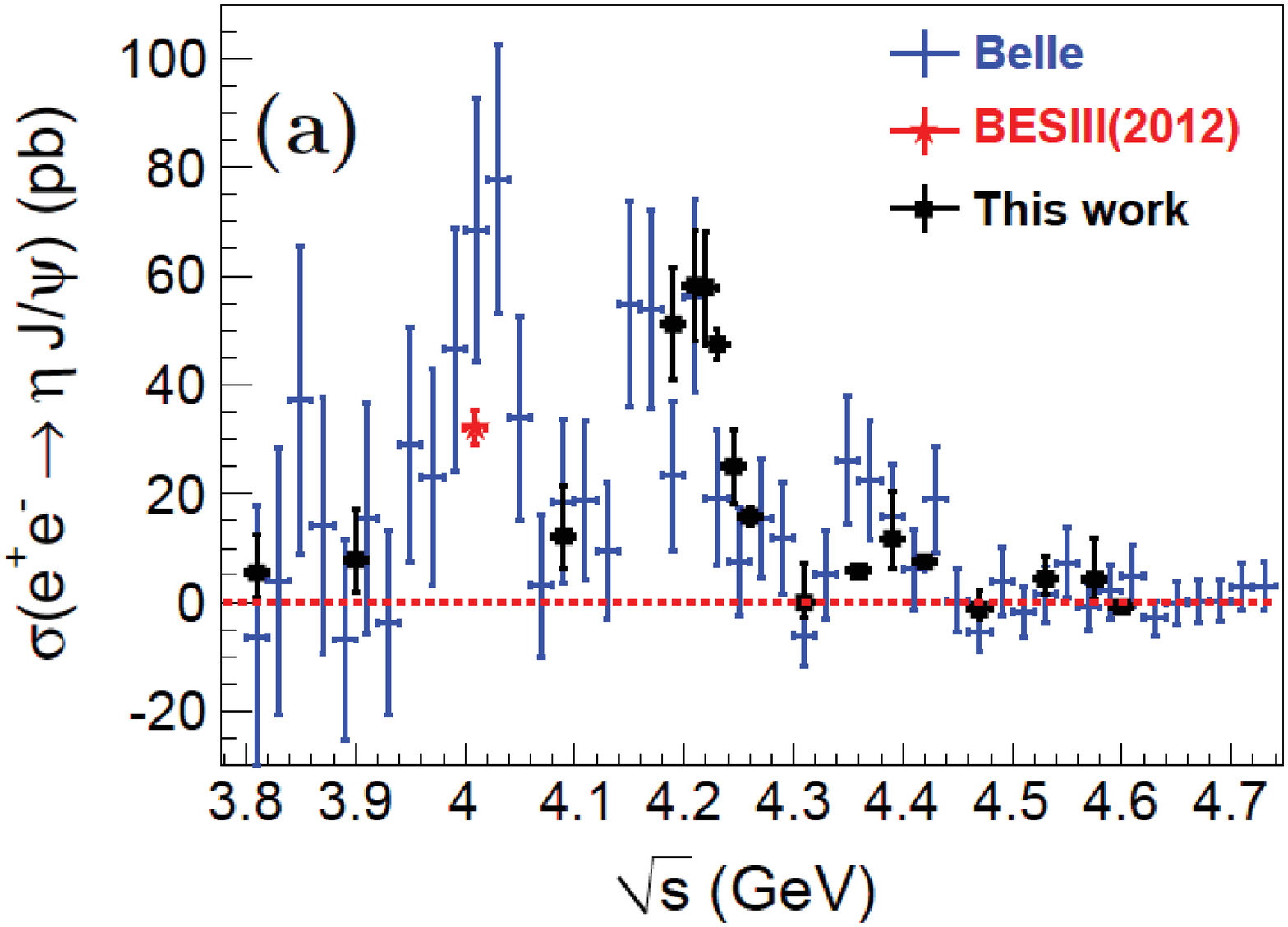}
\includegraphics[width=0.32\textwidth]{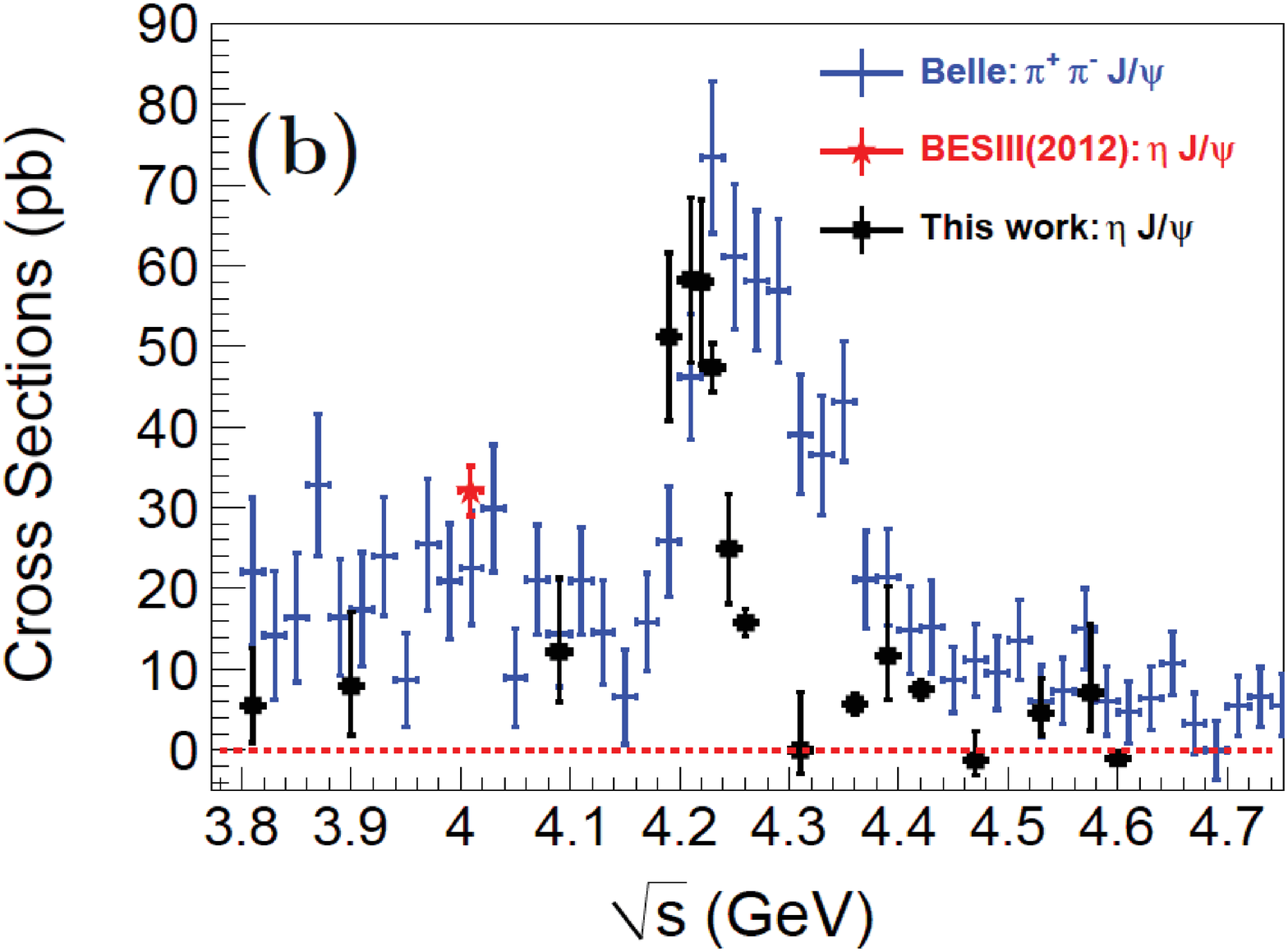}
\includegraphics[width=0.32\textwidth]{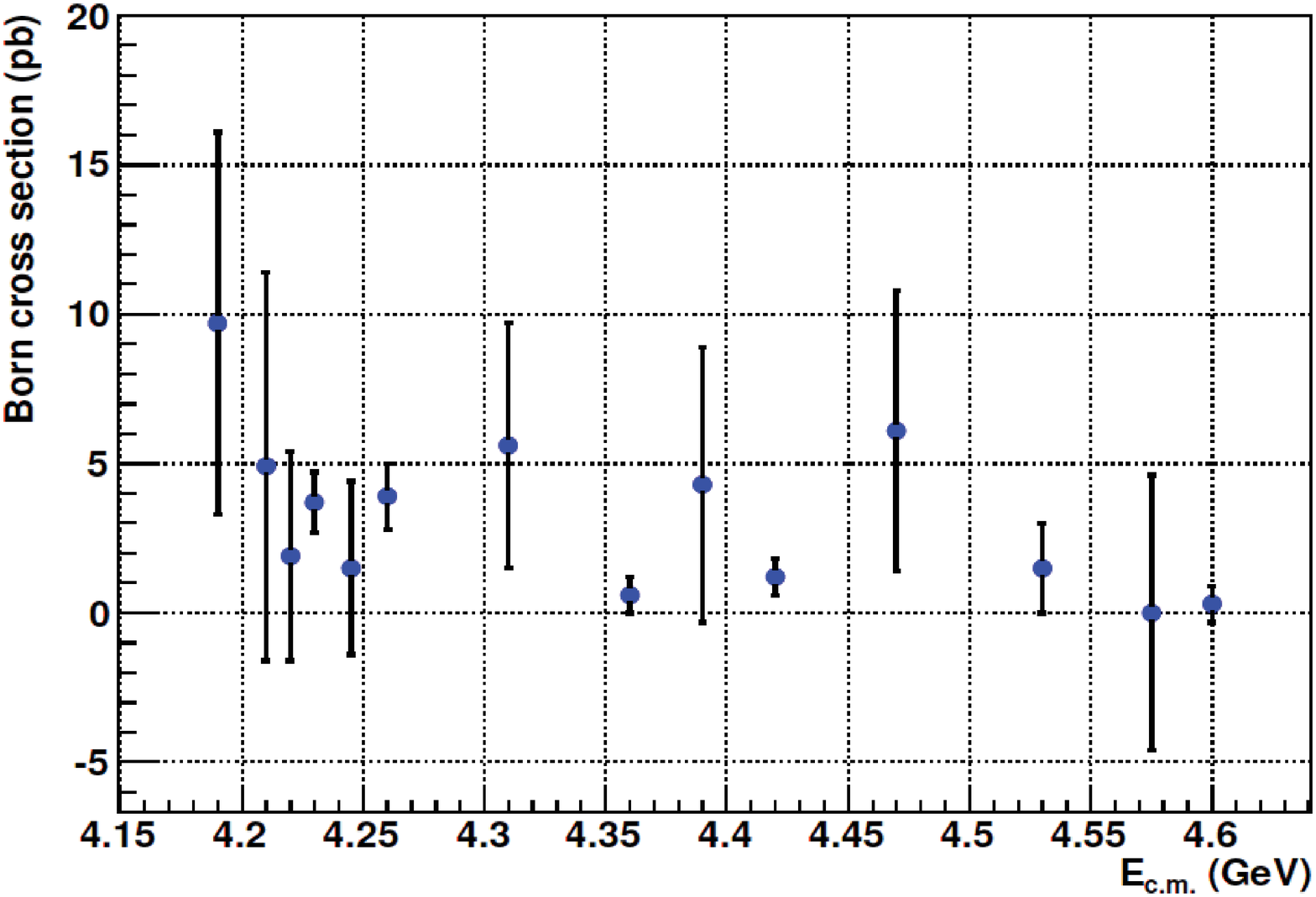}
\caption{A comparison of the measured Born cross section of $e^+e^-\rightarrow \eta J/\psi$ to (a) that of a previous measurement, (b) that of $e^+e^-\rightarrow\pi^+\pi^-J/\psi$ from Belle. (c) The Born cross sections of $e^+e^-\rightarrow \eta' J/\psi$.}
\label{fig:etajpsi}
\end{center}
\end{figure}

BESIII reported a new measurement of the Born cross sections of $e^+e^-\rightarrow \eta J/\psi$~\cite{etajpsi}, which are very consistent with the previous results~\cite{y25,y26} as shown in Figure~\ref{fig:etajpsi} (a).
Figure~\ref{fig:etajpsi} (b) indicates that the production mechanism of the $\eta J/\psi$ clearly differs from that of $\pi^+\pi^-J/\psi$. This could indicate the existence of a rich spectrum of $Y$ states in this energy region with different coupling strengths to the various decay modes.

The ratio of the Born cross section at 4.260\,GeV to that at 4.230\,GeV, $R(e^+e^-\rightarrow \eta J/\psi)$ = $\frac{\sigma^{4.260}(e^+e^-\rightarrow \eta J/\psi)}{\sigma^{4.230}(e^+e^-\rightarrow \eta J/\psi)}=0.33\pm0.04$, agrees very well with the ratio, $R(e^+e^-\rightarrow \omega\chi_{c0})$ = $\frac{\sigma^{4.260}(e^+e^-\rightarrow \omega\chi_{c0})}{\sigma^{4.230}(e^+e^-\rightarrow \omega\chi_{c0})}$ = 0.43$\pm$0.13. This may indicate that the production of $\eta J/\psi$ and $\omega\chi_{c0}$ are from the same source. More data around this energy region may be useful to clarify this interpretation.
\subsection{Cross sections of $e^+e^-\rightarrow \eta' J/\psi$}
BESIII observed significant $e^+e^-\rightarrow \eta' J/\psi$ 
signals at $\sqrt{s}$ = 4.230 and 4.260\,GeV, at the first time. And the corresponding Born cross sections are measured to be (3.7$\pm$0.7$\pm$0.3) and (3.9$\pm$0.8$\pm$0.3)\,pb, respectively. The upper limits of Born cross sections at the 90\% C.L. are set for the other 12 CM energy points as shown in Figure~\ref{fig:etajpsi} (c). Comparing with the measurement of $e^+e^-\rightarrow \eta J/\psi$, the Born cross section of $e^+e^-\rightarrow \eta' J/\psi$ is much smaller, which is in contradiction to the calculation in the framework of NRQCD~\cite{NRQCD}. There are two possible reasons contributing to the discrepancy. On one hand, the cross section of $e^+e^-\rightarrow \eta' J/\psi$ is investigated at order of $O(\alpha_s^4)$, therefore, higher order correction might need to be considered; on the other hand, gluonium component contributions may affect the results significantly, requiring the proportion of gluonic admixture in $\eta'$ to be further studied.

\begin{figure}[!htb]
\centering
\begin{overpic}[width=0.24\linewidth]{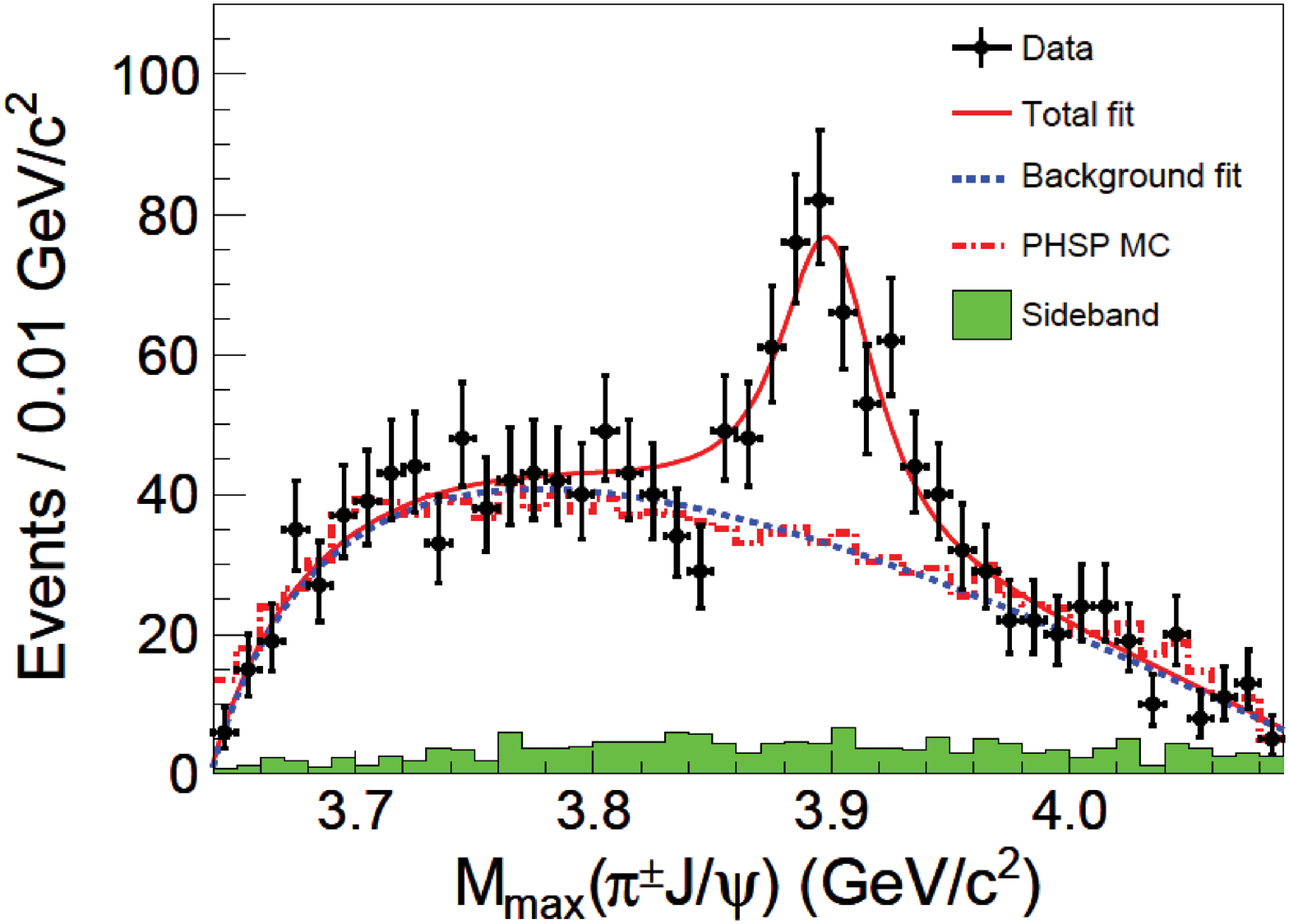}
\put(25,50){(a)}
\end{overpic}
\begin{overpic}[width=0.24\linewidth]{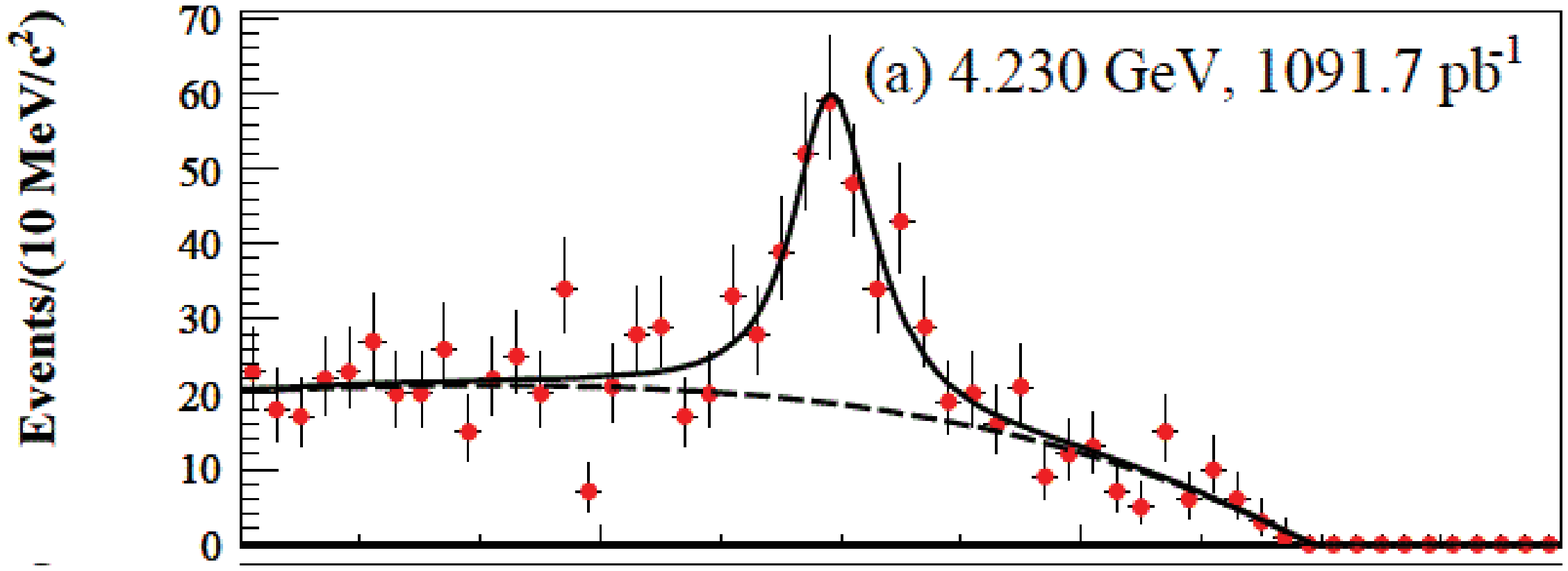}
\put(25,40){(b)}
\put(30,8){\tiny{M($\pi^0 J/\psi$)(GeV/c$^{2}$)}}
\end{overpic}
\begin{overpic}[width=0.24\linewidth]{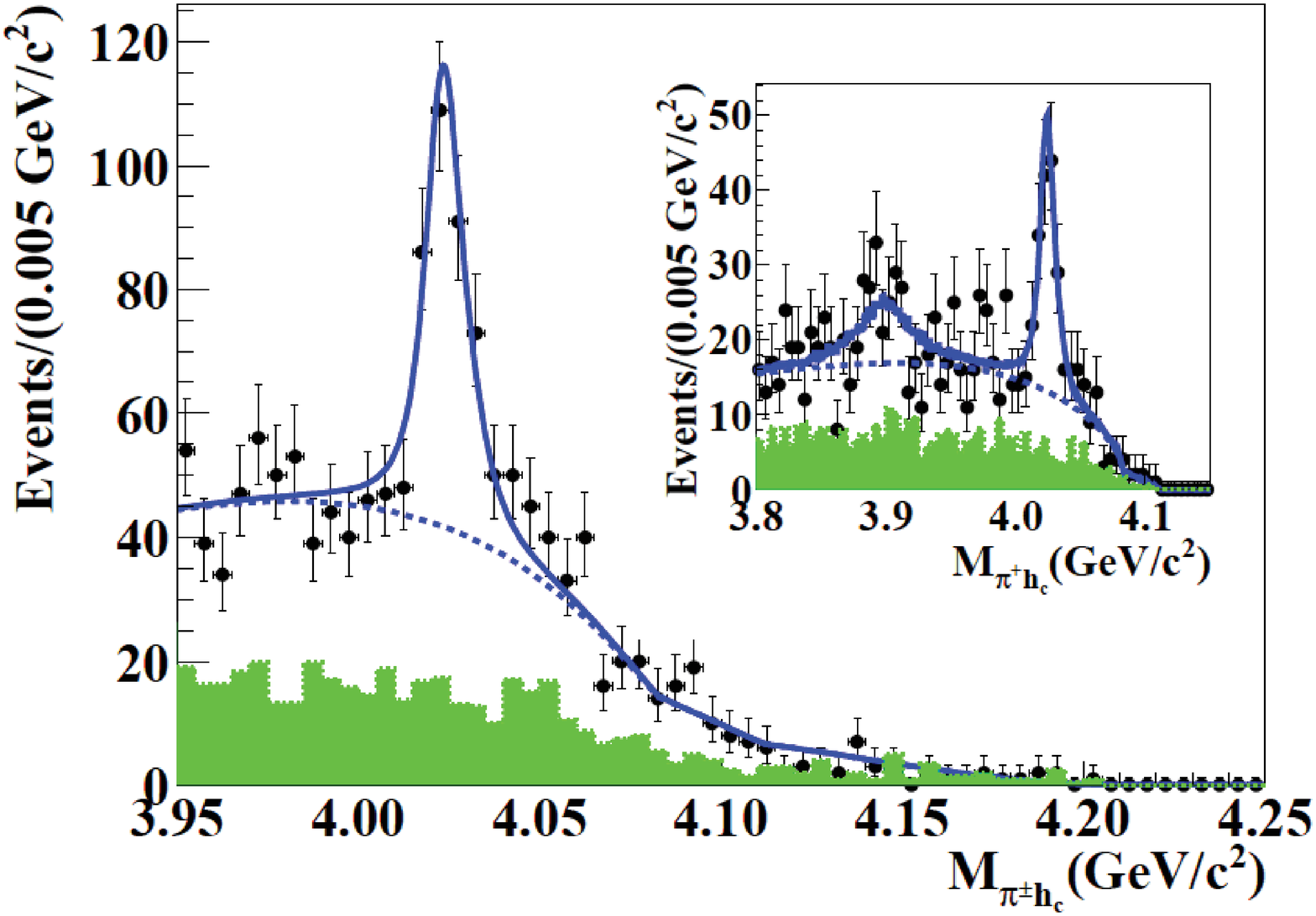}
\put(18,50){(c)}
\end{overpic}
\begin{overpic}[width=0.24\linewidth]{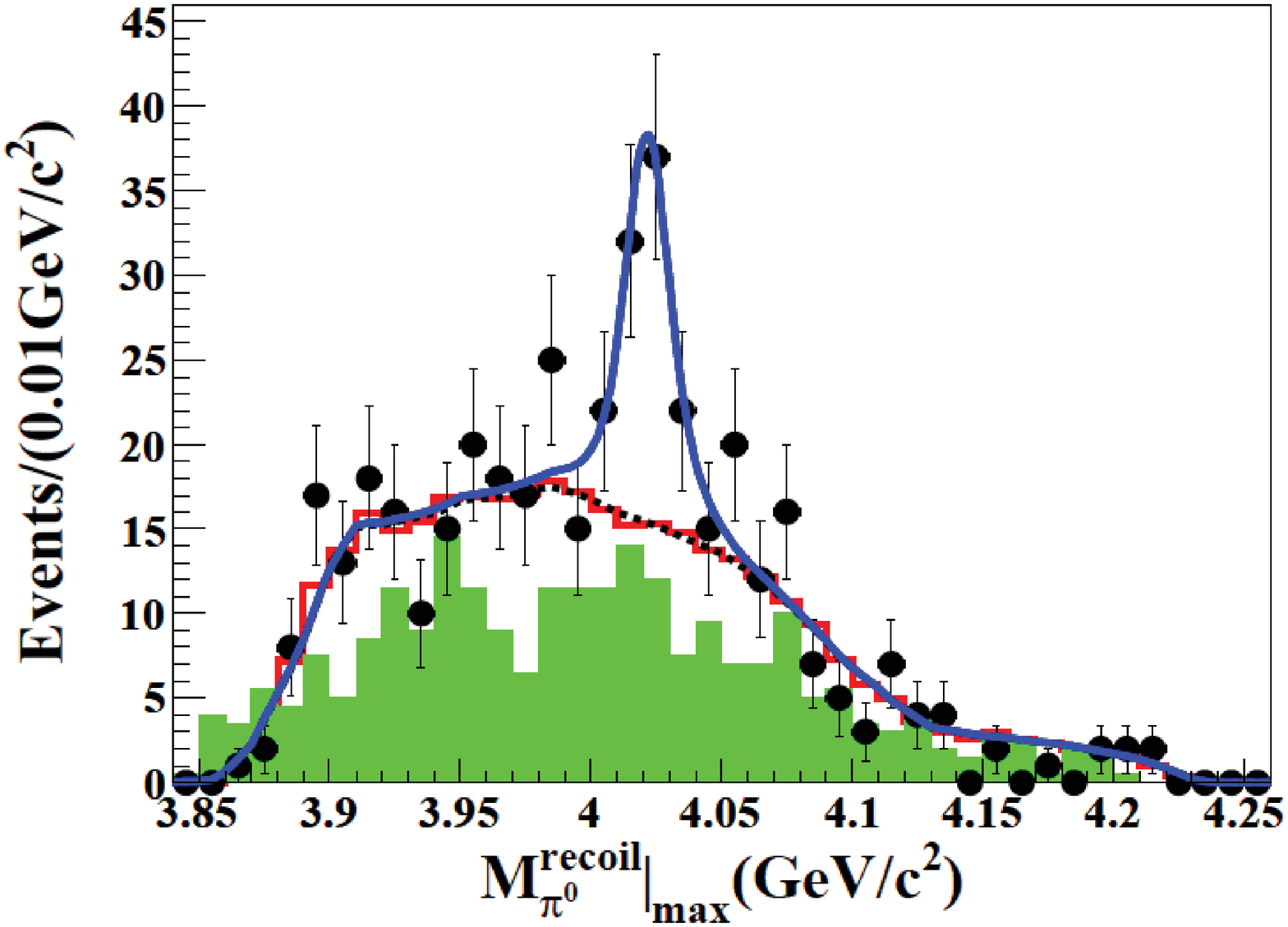}
\put(20,50){(d)}
\end{overpic}
\begin{overpic}[width=0.24\linewidth]{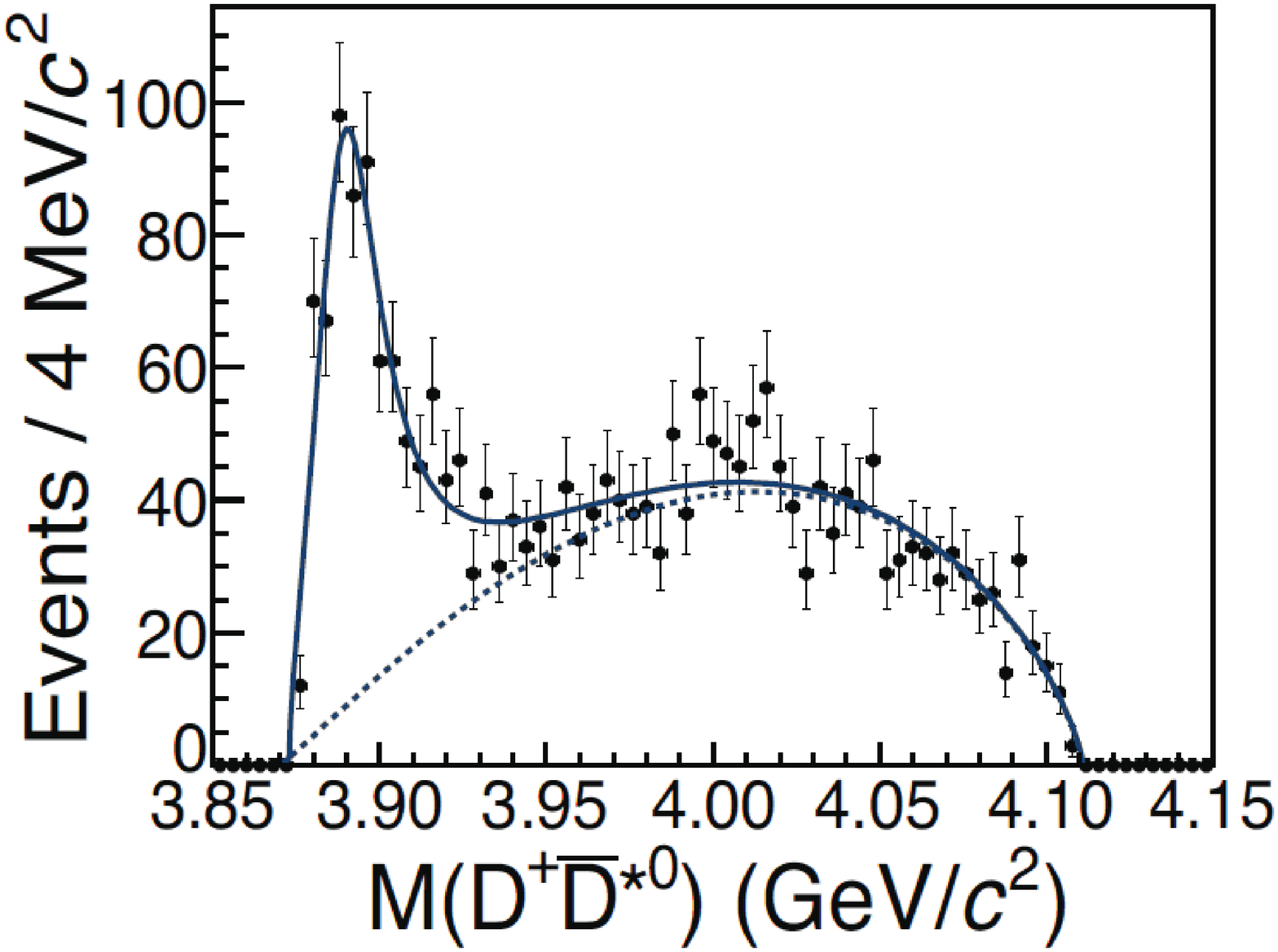}
\put(40,60){(e)}
\end{overpic}
\begin{overpic}[width=0.24\linewidth]{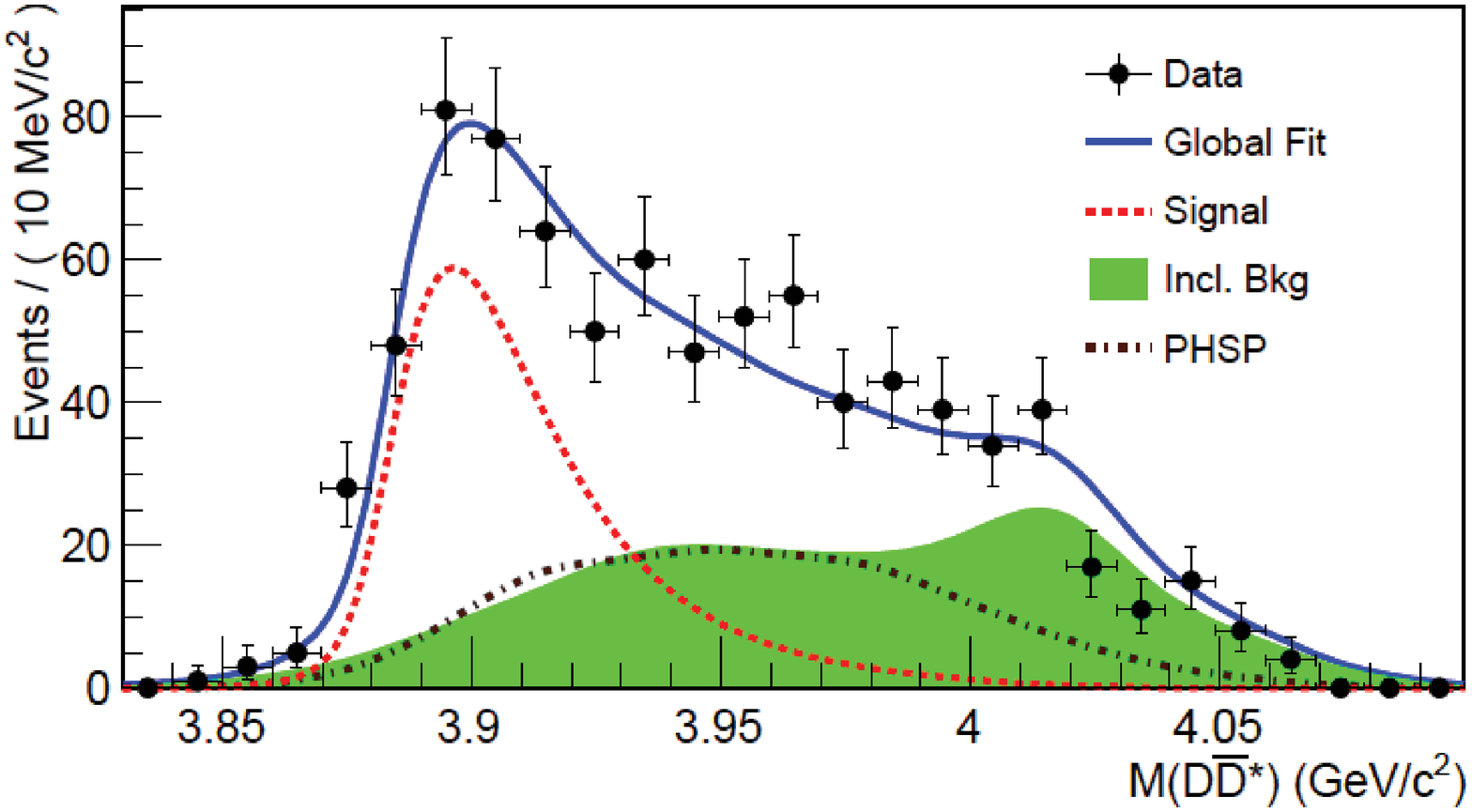}
\put(15,40){(f)}
\end{overpic}
\begin{overpic}[width=0.24\linewidth]{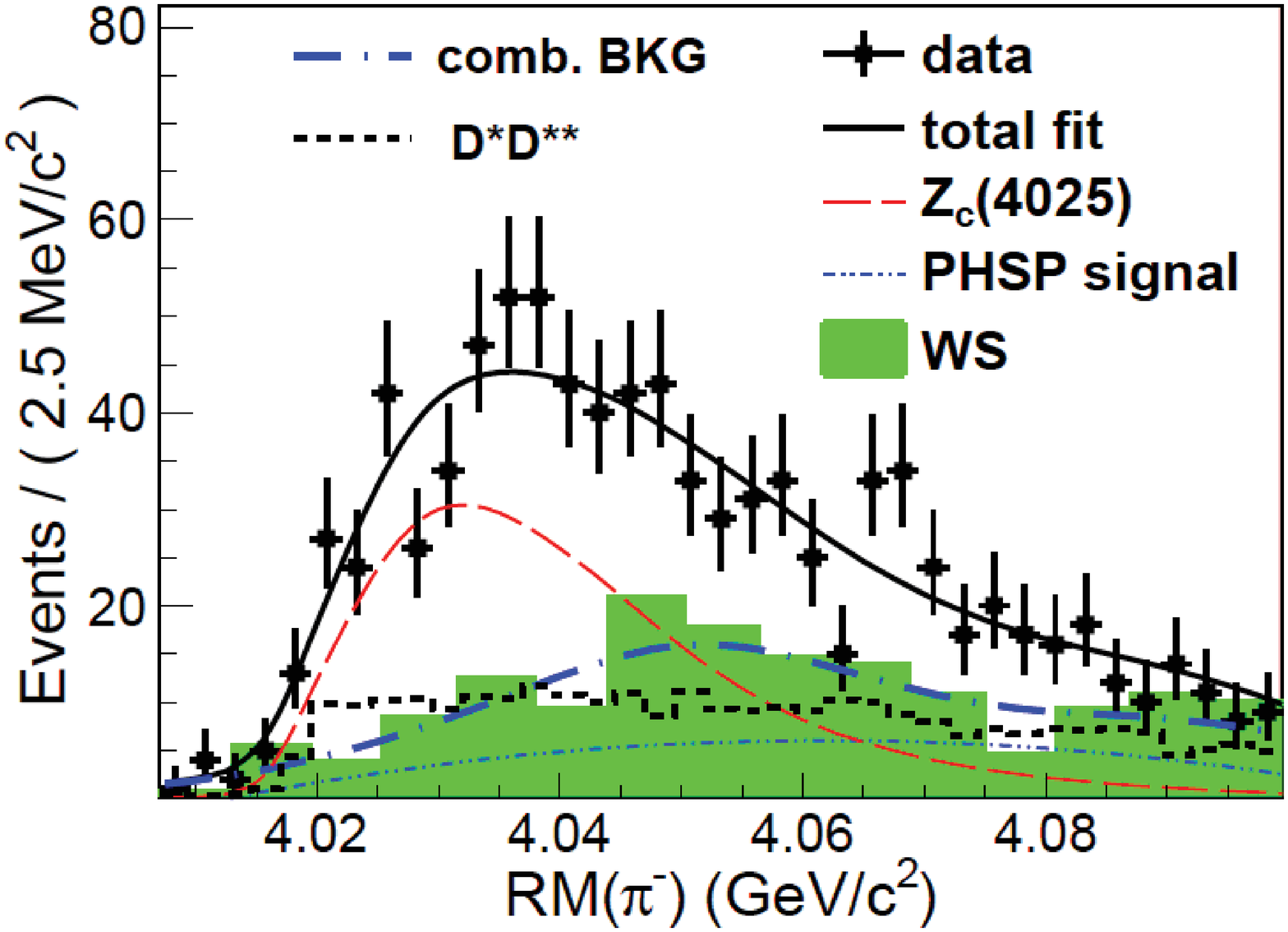}
\put(18,50){(g)}
\end{overpic}
\begin{overpic}[width=0.24\linewidth]{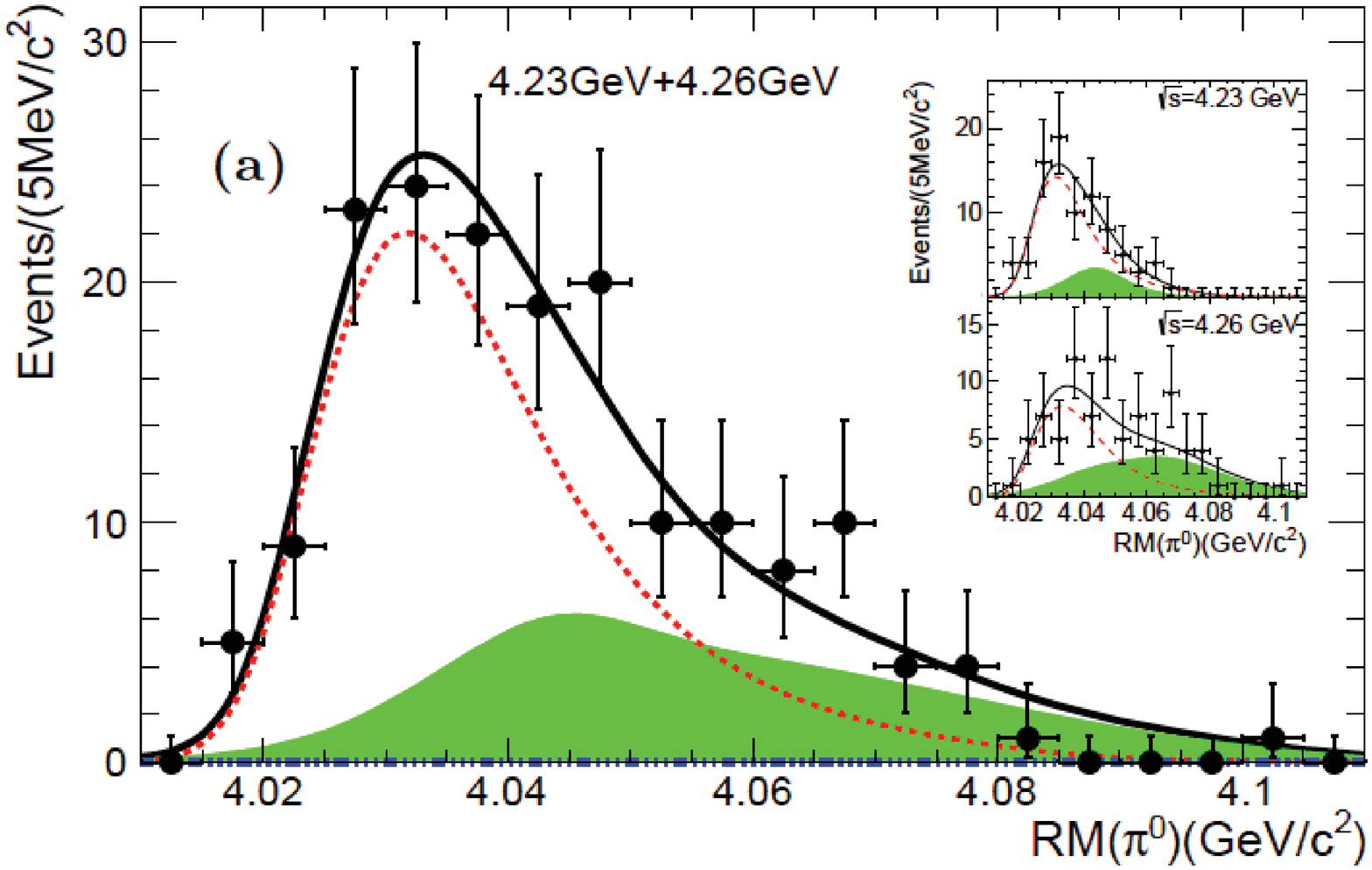}
\put(12,48){(h)}
\end{overpic}
\caption{Observed $Z_c$ states at BESIII.}
\label{fig:ZC}
\end{figure}

\section{$Z_c$}
Asides from the $Y$ states, several charged charmonium-like states: the $Z_c$(3900)$^{\pm}$, $Z_c$(3885)$^{\pm}$, $Z_c$(4020)$^{\pm}$, $Z_c$(4025)$^{\pm}$, as well as their isospin partners, the neutral states $Z_c$(3900)$^0$, $Z_c$(3885)$^0$, $Z_c$(4020)$^0$ and $Z_c$(4025)$^0$, were observed in the same mass region as the $Y$ states. This suggests that the nature of the $Y$ states could be related to that of the $Z_c$ states. Because the $Z_{c}^{\pm}$ couples to charmonium and has electric charge, it can not be a conventional $q\bar{q}$ meson, but must include at least two light quarks in addition to a $c\bar{c}$ pair. Proposed interpretations for $Z_{c}^{\pm}$ include hadronic molecules, hadro-quarkonia, tetraquark states, and kinematic effects. The precise structure of the $Z_{c}^{\pm}$ and other ``$XYZ$" states remains mysterious, and it is
clear that their further study will lead to a deeper understanding of the strong interaction in the non-perturbative regime.

Several $Z_c$ states have been observed via their decays into $\pi J/\psi$, $\pi h_c$, $D\bar{D}^*$ and $D^*\bar{D}^*$ at BESIII, as shown in Figure~\ref{fig:ZC}. The parameters of these $Z_c$ states are listed in Table~\ref{tab:Zc}. In addition, BESIII reports a search for $Z_c(3900)^{\pm}\rightarrow \omega\pi^{\pm}$~\cite{BESpipiomg} at $\sqrt{s}$ = 4.23 and 4.26\,GeV, and no $Z_c(3900)^{\pm}$ signal is observed in $e^+e^-\rightarrow \omega\pi^+\pi^-$.

\begin{table}[!htb]
\begin{center}
\begin{tabular}{c|c|c|l|l}
\hline\hline
State                              &  M (MeV/$c^2$) & $\Gamma$ (MeV) & Process & $\sqrt{s}$ (GeV)\\
\hline
(a) $Z_c(3900)^{\pm}$~\cite{3900+}  & $3899.0\pm3.6\pm4.9$ & $46\pm10\pm20$     & $e^+e^-\rightarrow\pi^+\pi^- J/\psi$   & 4.26 \\
(b) $Z_c(3900)^0$~\cite{pi0pi0jpsi}     & $3894.8\pm2.3\pm3.2$ & $29.6\pm8.2\pm8.2$ & $e^+e^-\rightarrow\pi^0\pi^0 J/\psi$   & 4.23,4.26,4.36 \\
(c) $Z_c(4020)^{\pm}$~\cite{pippimhc}  & $4022.9\pm0.8\pm2.7$ & $7.9\pm2.7\pm2.6$  & $e^+e^-\rightarrow\pi^{+}\pi^{-}h_{c}$ & 4.23,4.26,4.36 \\
(d) $Z_c(4020)^0$~\cite{pi0pi0hc}     & $4023.9\pm2.2\pm3.8$ & fixed to $\Gamma(Z_c(4020)^{\pm})$ & $e^+e^-\rightarrow\pi^{0}\pi^{0}h_{c}$ & 4.23,4.26,4.36\\
(e) $Z_c(3885)^{\pm}$~\cite{3885+}  & $3883.9\pm1.5\pm4.2$ & $24.8\pm3.3\pm11.0$ & $e^+e^-\rightarrow \pi^{\pm} (D\bar{D}^*)^{\mp}$  & 4.26 \\
(f) $Z_c(3885)^{0}$~\cite{3885-0}   & $3885.7^{+4.3}_{-5.7}\pm8.4$ & $35^{+11}_{-12}\pm15$ & $e^+e^-\rightarrow \pi^{0} (D\bar{D}^*)^0$ & 4.23,4.26\\
(g) $Z_c(4025)^{\pm}$~\cite{4025+}  & $4026.3\pm2.6\pm3.7$ & $24.8\pm5.6\pm7.7$ & $e^+e^-\rightarrow \pi^{\pm}(D^*\bar{D}^*)^{\mp}$ & 4.26\\
(h) $Z_c(4025)^{0}$~\cite{4025-0}   & $4025.5^{+2.0}_{-4.7}\pm3.1$ & $23.0\pm6.0\pm1.0$ & $e^+e^-\rightarrow \pi^0 (D^*\bar{D}^*)^0$ & 4.23,4.26\\
\hline\hline
\end{tabular}
\caption{Observed $Z_c$ states at BESIII.}
\label{tab:Zc}
\end{center}
\end{table}

\section{Conclusion}
Some significant progress in charmonium-like studies at BESIII have been presented. A number of transitions between different charmonium-like states are observed, and more studies are needed to make clear their relations. The nature of the charmonium-like states are still mysterious, and some expected states and decay modes are missing. BESIII will collect more data for $XYZ$ study. With some analysis are on going, more results will come up soon.

\end{document}